\documentclass[a4paper,twocolumn,11pt,accepted=2026-02-11]{quantumarticle}
\pdfoutput=1
\usepackage[utf8]{inputenc}
\usepackage[english]{babel}
\usepackage[T1]{fontenc}
\usepackage{stmaryrd}
\usepackage{graphicx}
\usepackage{bm}
\usepackage{subcaption}
\usepackage{hyperref}

\usepackage{amsmath}
\usepackage{amsthm}
\usepackage{amsfonts}
\usepackage{xcolor}
\usepackage{algorithm2e}
\usepackage{braket}
\usepackage{float}
\usepackage{adjustbox}
\usepackage{array}

\begin{document}
\title{Coprime Bivariate Bicycle Codes and Their Layouts on Cold Atoms}

\author{Ming Wang}
\email{mwang42@ncsu.edu}
\affiliation{Department of Computer Science,\\
North Carolina State University,\\
Raleigh,
North Carolina, USA
}

\author{Frank Mueller}%
 \email{fmuelle@ncsu.edu}
\affiliation{Department of Computer Science,\\
North Carolina State University,\\
Raleigh,
North Carolina, USA
}

\begin{abstract}
  Quantum computing is deemed to require error correction at scale to
  mitigate physical noise by reducing it to lower noise levels while
  operating on encoded logical qubits. Popular quantum error
  correction schemes include CSS code, of which surface codes provide
  regular mappings onto 2D planes suitable for contemporary quantum
  devices together with known transversal logical gates. Recently,
  qLDPC codes have been proposed as a means to provide denser encoding
  with the class of bivariate bicycle (BB) codes promising feasible
  design for devices.
  
  This work contributes a novel subclass of BB codes suitable for
  quantum error correction. This subclass employs {\em coprimes} and
  the product $xy$ of the two generating variables $x$ and $y$ to
  construct polynomials, rather than using $x$ and $y$ separately as
  in vanilla BB codes. In contrast to vanilla BB codes, where
  parameters remain unknown prior to code discovery, the rate of the
  proposed code can be determined beforehand by specifying a factor
  polynomial as an input to the numerical search algorithm.  Using
  this coprime-BB construction, we found a number of surprisingly
  short to medium-length codes that were previously unknown. We also
  propose a layout on cold atom arrays tailored for coprime-BB
  codes. The proposed layout reduces both move time for short to
  medium-length codes and the number of moves of atoms to perform
  syndrome extractions. We consider an error model with global laser
  noise on cold atoms, and simulations show that our proposed layout
  achieves significant improvements over prior work across the
  simulated codes.
\end{abstract}

\maketitle


\section{Introduction}
Quantum information is susceptible to errors during storage and
operation. As the number of qubits in a quantum circuit increases, so
does the frequency of errors. Therefore, quantum error correction
(QEC) is a cornerstone of advancing from the current noisy
intermediate-scale quantum (NISQ) era to the next era of
fault-tolerant quantum (FTQC) computing. Typically, a QEC code is
characterized by a 3-tuple $\llbracket n, k, d\rrbracket $, indicating
that the code utilizes $n$ physical qubits to encode $k$ logical
qubits and can correct up to $\lfloor (d-1)/2 \rfloor/$ errors. This
encoding process incurs overhead, quantified by
the code rate $k/n$.  Among the various QEC codes, quantum Low-Density
Parity-Check (qLDPC) codes stand out~\cite{mackay04, Tillich14,
  10.1145/3519935.3520017,Breuckmann2021qldpc, Bravyi_2024} due to
their lower-weight stabilizers, low overhead, and high thresholds. As
a special case of qLDPC codes, the surface codes, which also feature
low-weight stabilizers, are the most commonly used codes in quantum
computing due to their simple 2D grid structure and simplicity to
perform logical operations~\cite{PhysRevA.86.032324}. However, surface
codes require significant overhead. For example, the rotated surface
code has parameters $\llbracket L^2, 1, L\rrbracket $, meaning that it
requires $L^2$ physical qubits to protect one logical qubit.

In contrast, research has demonstrated the existence of ``good'' qLDPC
codes~\cite{10.1145/3519935.3520017} indicating that qLDPC codes can
have $k$ and $d$ scaling linearly as $n$ grows, i.e. codes with
parameters $\llbracket n, k=\Theta(n),
d=\Theta(n)\rrbracket$. However, having asymptotically good LDPC codes
does not necessarily mean having better parameters than codes designed
for short to medium lengths. Moreover, the structure of codes can
greatly influence the design of hardware and vice versa. Therefore,
for practical purposes, people started to seek qLDPC codes that are
easy to implement on hardware and have good finite-length
performance. For example, Panteleev and
Kalachev~\cite{panteleev2021degenerate} proposed generalized bicycle
(GB) codes along with the BP-OSD decoder that focus on medium-length
performance. It is noteworthy that they also discovered the
$\llbracket 126, 12, 10\rrbracket$ GB code, which has the same
parameters as the code we will analyze later. Another similar qLDPC
code, the BB code~\cite{Bravyi_2024}, has received much attention
because they have high thresholds, toric layout, and can be embedded
on two planes. Benefiting from the quasi-cyclic and two-thickness
properties of BB codes, recent papers have established the feasibility
of BB codes on different architectures, including cold atom
arrays~\cite{viszlai2023matching, poole2024architecture}, trapped
ion\cite{PhysRevLett.133.180601}, and
superconducting~\cite{berthusen2024toward, Bravyi_2024}. Due to the
high connectivity of qLDPC codes, they are considered hard to
implement on a superconducting-based platform unless a multi-layer
architecture is used~\cite{berthusen2024toward}. Benefiting from their
all-to-all connectivity, cold atom array and trapped ion systems have
certain advantages in implementing qLDPC codes. In particular, Viszlai
et al.~\cite{viszlai2023matching} showed that the quasi-cyclic
structure of BB codes can be easily implemented using 2D atom array
acousto-optic deflectors (AOD) for atom movement.

Constructing BB codes is, nonetheless, a time-consuming process as one
has to search for combinations of polynomials to construct a
code. Moreover, the parameters of codes can not be guaranteed. This
challenge motivated us to develop an algorithm that accelerates the
search for good BB codes and even constructs codes with pre-determined
parameters. Our work differentiates itself from these approaches by
introducing a general and efficient algorithm for searching BB codes
in the form described in Eq.~(\ref{ibm_form}), which is also the form
of codes used by related works. Furthermore, we proposed a novel
algorithm that allows us to search for a subclass of BB codes with the
desired dimensions, which we call coprime-BB codes. Coprime-BB codes
generalize the form in Eq.~(\ref{ibm_form}) by allowing mixed terms
but also have a restriction of using coprime circulant matrix sizes,
making them unattainable through searches limited to the polynomial
form of this equation.  Thus, coprime-BB codes provide more flexibility
for different scenarios on top of the originally proposed BB codes.

To summarize, this work makes the following contributions:
\begin{itemize}
\item We propose a fast numerical algorithm to search for good BB
  codes by excluding certain polynomial combinations.
\item We propose a new construction of BB codes that
  allows us to customize the code rate {\em before} performing a
  search, much in contrast to prior search techniques that identified
  the rate only after returning a new code as a search result. This
  new method involves selecting two coprime numbers and a factor
  polynomial, leading us to name this subclass {\em coprime-BB codes}.
\item We study the properties of coprime-BB codes, devise a novel
  method to reduce the search space for unknown codes and develop a
  new layout tailored for quantum devices utilizing cold atom
  arrays. The simulations show that the proposed layout achieves
  a lower logical error rate than under layouts of prior
  work~\cite{viszlai2023matching}. This layout benefits from the
  coprime properties of our codes, which the traditional layout does
  not exploit, i.e., our coprime codes result in superior performance.
\end{itemize}

\section{Background}
\subsection{Calderbank-Shor-Steane(CSS) codes}
Stabilizer codes are among the most commonly used codes in quantum
error correction. One can measure each stabilizer to infer both the
type and location of errors in a multi-qubit system. To construct such
a code, all stabilizers must commute with each other. Thus, they have
a common eigenspace and form a stabilizer group $\mathcal{S}$. The
code space defined by such group is
\begin{equation}
    \mathcal{C}=\{\ket{\psi} |\: s\ket{\psi}=\ket{\psi},\: \forall s\in \mathcal{S}\}.
\end{equation}

An $\llbracket n, k, d\rrbracket $ stabilizer code can be defined by
$n-k$ independent stabilizers, allowing us to encode $k$ qubits of
logical information into an $n$-qubit block tolerating up to
${\lfloor (d-1)/2 \rfloor}$ errors. CSS codes are an important class
of stabilizer with two sets of stabilizers, $X$-type and $Z$-type,
represented by parity-check matrices $H_X$ and $H_Z$,
respectively. Each row in a parity-check matrix corresponds to one
stabilizer, and each column corresponds to a qubit. A ``1'' entry
signifies an $X$ or $Z$ operator (depending on whether it is in $H_X$
or $H_Z$), while a ``0'' indicates the identity. Consequently, a
$X$-type stabilizer acts as $X$ or the identity on each qubit, and a
$Z$-type stabilizer acts as $Z$ or the identity on each qubit. Errors
can therefore be corrected by handling $Z$ errors and $X$ errors
separately. Since all stabilizers must commute with each other, it
follows directly that for a CSS code $H_XH_Z^T=0$.

\subsection{Bivariate Bicycle Codes}
BB codes~\cite{Bravyi_2024} are a class of CSS codes and LDPC
codes. In this context, LDPC means that the stabilizers have bounded
weight ensuring low density in both rows and columns of the
parity-check matrices.

Let $S_m$ be the shift matrix of
size $m$, defined as
\begin{equation}
    S_m=I_m >> 1,
\end{equation}
where ``$>>$'' denotes the right cyclic shift for each row, and $I_m$ is the $m\times m$ identity matrix. For example, 
\begin{equation}
    S_3 = \begin{bmatrix}
        0&1&0\\
        0&0&1\\
        1&0&0
    \end{bmatrix}.
\end{equation}
By defining $x=S_l\otimes I_m$ and $y=I_l\otimes S_m$, it is easy to
verify that $xy=yx$ using the mixed-product property of the Kronecker
product. This definition forms a bijection from the set of monomials
$\{x^iy^j | 0\leq i<l, 0\leq j <m\}$ to the set of $(lm)\times (lm)$
matrices generated by $x$ and $y$. Therefore, we can interchangeably
use polynomials or monomials in $x$ and $y$ to represent their
corresponding matrices. It is also straightforward that
$x^l=y^m=I$. The BB codes can be defined by two polynomials,
$A = a(x,y)$ and $B = b(x,y)$, and the parity check matrices for BB
code are defined as
\begin{equation}
    \begin{split}
        H_X&=[A|B]\\
        H_Z&=[B^T|A^T].
    \end{split}
\end{equation}



and it meets the CSS condition as $H_X H_Z^T=AB+BA=2AB=\bm{0}$ in
$\mathbb{F}_2$ since $x$ and $y$ commute. In~\cite{Bravyi_2024}, the
authors restricted the polynomials to shapes of
\begin{equation}
    \label{ibm_form}
    \begin{split}
    a(x,y) &= x^a+y^b+y^c\\
    b(x,y) &= y^d+x^e+x^f.
    \end{split}
\end{equation}
Therefore, we can write $A$ and $B$ as $A=A_1+A_2+A_3$ and
$B=B_1+B_2+B_3$. Each polynomial has three terms, making each
stabilizer supported by six qubits.  Besides, we know that
$A^T=A_1^T+A_2^T+A_3^T=A_1^{-1}+A_2^{-1}+A_3^{-1}$ as $A_i$ is the
power of $x$ or $y$, which are permutation matrices. 
Similarly, we have
$B^T=B_1^{-1}+B_2^{-1}+B_3^{-1}$. It is well
known~\cite{PhysRevA.54.1098} that for any CSS code,
$k=n-\text{rank}(H_X)-\text{rank}(H_Z)$. For BB codes, according
to Lemma 1 in~\cite{Bravyi_2024}, $\text{rank}(H_X)=\text{rank}(H_Z)$,
so this expression simplifies to
\begin{equation}
\label{bb_dim}
    k=2lm- 2\text{rank}(H_X)=2lm- 2\text{rank}(H_Z).
\end{equation}
\subsection{Cold Atom Arrays}
In cold array-based quantum computers, logical qubits are encoded in a
two-dimensional physical atom array, allowing for high-fidelity
single- and two-qubit operations to be performed in
parallel~\cite{bluvstein2024logical}. Specifically, atoms are loaded
into optical traps generated by a spatial light modulator (SLM) and
acousto-optic deflectors (AODs). The AOD-based traps enable both
vertical and horizontal movement of qubits, thereby providing
arbitrary connectivity.  A global qubit rotation can be carried out in
parallel by illuminating the entire array with Raman
excitation. Two-qubit gates can also be executed in parallel by
collectively moving qubits. After a global Rydberg laser pulse is
applied, entangling gates are performed on pairs of qubits that are
brought sufficiently close to each other (closer than the Rydberg
blockade radius). The atom array in AOD traps can be moved
horizontally, vertically, or even stretched~\cite{bochenqubit}, which
facilitates the syndrome extraction for BB codes.

\begin{figure}[!t]
  \centering
  \includegraphics[width=3.3 in]{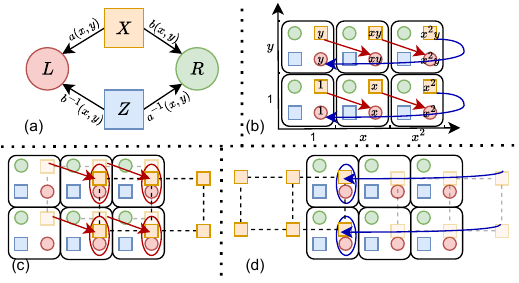}
  \caption{%
    (a) Connections defined by the polynomials \( a(x, y) \) and \(
    b(x, y) \) between \( X \), \( Z \), \( L \), and \( R \) qubits.
    (b) The layout from~\cite{viszlai2023matching} for a BB code with
    \( l = 3 \) and \( m = 2 \). Arrows indicate the CNOT pairs to be
    performed when \( a(x, y) = x \). In this case, CNOTs are applied
    between the following \( X \) ancilla and \( L \) data qubit
    pairs:
    \( (X_1, L_x) \), 
    \( (X_y, L_{xy}) \), 
    \( (X_x, L_{x^2}) \), 
    \( (X_{xy}, L_{x^2y}) \), 
    \( (X_{x^2}, L_1) \), 
    and \( (X_{x^2y}, L_y) \).
    (c--d) Lighter orange squares indicate the initial positions of
    the \( X \) checks before movement. Each check monomial requires
    two moves.
    (c) Rightward movement enables CNOTs between \( (X_1, L_x) \), \(
    (X_y, L_{xy}) \), \( (X_x, L_{x^2}) \), and \( (X_{xy}, L_{x^2y})
    \).
    (d) Leftward movement enables CNOTs between \( (X_{x^2}, L_1) \)
    and \( (X_{x^2y}, L_y) \).
  }
  \label{bb_layout}
\end{figure}

In~\cite{viszlai2023matching}, Viszlai et al. propose a layout for BB
codes on atom arrays that leverages the quasi-cyclic property of these
codes. We will refer to this layout as {\em ``BB layout''} for the
rest of paper; As shown in Eq.~(\ref{bbcodes}), one can split the data
qubits into two blocks, $L$ and $R$, corresponding to the first $lm$
columns and last $lm$ columns of $H_X$ and $H_Z$. The ancilla qubits
are divided into $X$ and $Z$ blocks corresponding to the rows of $H_X$
and $H_Z$, respectively. Each qubit can be addressed using a letter
from $X,Z,L,R$ with a monomial in $\{x^iy^j|0\leq i<l, 0\leq
j<m\}$. This labeling facilitates the process to identify the supports
of an ancilla qubit: simply multiply the polynomial associated with
the qubit by the relevant monomial by identifying $x^l=y^m=1$. In
other words,
\begin{equation}
    x^ay^b\cdot x^cy^d=x^{(a+c)\text{mod } l}y^{(b+d)\text{mod } m}.
\end{equation} 
For example, given $a(x,y)=1+x$ and $b(x,y)=1+y$, we immediately know
that the ancilla qubit $X_{xy}$ supports $L_{xy}, L_{x^2y}, R_{xy}$
and $R_{xy^2}$. In addition, we can perform syndrome extraction in
parallel by mapping qubits with the same labels, i.e.,
$X_{x^ay^b}, Z_{x^ay^b}, L_{x^ay^b}, R_{x^ay^b}$, to a subgrid with
two-dimensional coordinates $(a,b)$, as illustrated in
Fig.~\ref{bb_layout}. This layout allows any check to be executed by
moving ancilla qubits cyclically along the vertical and/or horizontal
directions. Qubits near the boundary wrap around to the opposite side,
which may require multiple moves to complete the check. For example,
if $a(x,y)=x$, we need two steps to complete the syndrome extraction 
specified by $a(x,y)$ as illustrated in Fig.\ref{bb_layout} (c-d).

\SetKwComment{Comment}{/* }{ */}
\RestyleAlgo{ruled}

\section{Proposed Search Algorithms}
\subsection{Code equivalence}
In~\cite{Bravyi_2024}, BB codes are obtained through a numerical
search. To accelerate this search, we reduce the search space by
eliminating some codes with the same $n,k,d$ parameters. 

{\em Equivalence of BB codes ---}It is straightforward to prove that
the following four codes
\begin{equation}
    \begin{split}
        \mathcal{C}_1: H_X&=[A|B], H_Z=[B^T|A^T]\\
        \mathcal{C}_2: H_X&=[A^T|B^T], H_Z=[B|A]\\
        \mathcal{C}_3: H_X&=[B|A], H_Z=[A^T|B^T]\\ 
        \mathcal{C}_4: H_X&=[B^T|A^T], H_Z=[A|B]
    \end{split}
    \label{eqiv}
\end{equation}
share the same parameters, allowing us to search within only one class
of these codes. These equivalences can also be viewed as special cases of Theorem 6 in \cite{PhysRevA.109.022407}.

\textbf{Proof:} According to Lemma 1 in~\cite{Bravyi_2024}, every
bivariate bicycle code has the same distance over $X$ or $Z$ and
$\text{rank}(H_X) = \text{rank}(H_Z)$. Let us assume the distances and
dimensions of $\mathcal{C}_1, \dots, \mathcal{C}_4$ are
$d_1, \dots, d_4$ and $k_1,\dots, k_4$. For CSS codes,
$k=n-\text{rank}(H_X)-\text{rank}(H_Z)$ and it is easy to see that
$k_1=k_4$ and $k_2=k_3$ since one can get one code from another by
swapping $H_X$ and $H_Z$.  Thus, it is sufficient to prove that
\begin{itemize}
    \item $d_1=d_4$ and $d_2=d_3$.
    \item $k_1=k_2$ and $d_1=d_2$.
\end{itemize}
to ensure $\mathcal{C}_1,\dots, \mathcal{C}_4$ has the same parameters.

Let $\bm{f}$ be the binary string of an arbitrary logical $X$ operator
of $\mathcal{C}_1$, where $f_i=1$ if the operator acts as $X$ on qubit
$i$ and $f_i=0$ if it acts as $I$. By definition, logical operators
commute with every $Z$ stabilizer, thus satisfying
\begin{equation}
\label{z_commute}
    [B^T|A^T]\bm{f}^T=\bm{0}.
\end{equation}
For the sake of simplicity, given a binary string $\bm{x}$, we use
$Z^{\bm{x}}$ to represent a Pauli string
$Z^{x_1}\otimes Z^{x_2}\otimes\cdots \otimes Z^{x_n}$. Therefore,
Eq.~(\ref{z_commute}) indicates that $Z^{\bm{f}}$ is a logical $Z$
operator of $\mathcal{C}_4$ as $[B^T|A^T]$ is the $H_X$ of
$\mathcal{C}_4$. Hence, any $X$ operator of $\mathcal{C}_1$ is a $Z$
operator of $\mathcal{C}_4$, i.e., $d_1\geq d_4$. Similarly,
$d_4\geq d_1$. Therefore, $d_4=d_1$. Using the same reasoning, we can
prove that $\mathcal{C}_2$ and $\mathcal{C}_3$ have the same distance.

Next, we will prove that $\mathcal{C}_1$ and $\mathcal{C}_2$ have the
same distance. Let $C_l$ be the anti-diagonal matrix of size
$l\times l$. We know $C_lC_l=I$ and $C_lMC_l=M^T$ for any given matrix
$M$ of the same size. As $C_l$ is a full-rank matrix and
\begin{equation}
    [A^T|B^T] = C_{lm} [A|B] {\begin{bmatrix}
        C_{lm}&0\\
        0&C_{lm}
    \end{bmatrix},}
\end{equation}
we know $[A^T|B^T] $ and $ [A|B]$ have the same rank and
$k_1=k_2$. Let $Z^{\bm{p}}$ be an arbitrary logical $Z$ operator of
$C_2$, where $\bm{p}=(\bm{\alpha}|\bm{\beta})$ is a length $n$ binary
vector and $\bm{\alpha},\bm{\beta}$ are binary vectors of length
$n/2$. By definition, we have
$[A^T|B^T]\bm{p}^T=A\bm{\alpha}^T+B\bm{\beta}^T=\bm{0}$, i.e.,
\begin{equation}
    C_{lm}AC_{lm}\bm{\alpha}^T+C_{lm}BC_{lm}\bm{\beta}^T=\bm{0}.
\end{equation}
Recall that $C_{lm}C_{lm}=I$. By multiplying both sides with $C_{lm}$,
we get
\begin{equation}
    AC_{lm}\bm{\alpha}^T+BC_{lm}\bm{\beta}^T=\bm{0},
\end{equation}
and we know that $(\bm{\alpha} C_{lm}^T|\bm{\beta} C_{lm}^T)$ is a
logical $Z$ operator of $\mathcal{C}_1$ with the same weight of
$\bm{p}$. Therefore, $d_1\leq d_2$ and, similarly, $d_2\leq
d_1$. Thus, $d_2=d_1$.\qed

We note that these two codes,
$\mathcal{C}_1: H_X=[A|B], H_Z=[B^T|A^T]$ and
$\mathcal{C}_5: H_X=[A^T|B], H_Z=[B^T|A]$, do not always have the same
parameters. For example, when $l=6, m=12$, the code constructed by
$a(x,y)=x^4+y^2+y^6$ and $b(x,y)=y^5+x^3+x^4$ is a
$\llbracket 144,8,10\rrbracket $ code whereas the code constructed by
$a(x,y)=x^2+y^6+y^{10}$ and $b(x,y)=y^5+x^3+x^4$ is a
$\llbracket 144,8,8\rrbracket $ code.

\subsection{Searching for BB Codes}
Based on the equivalence, the accelerated search algorithm is
described in Algorithm~\ref{alg:bb}. The function
\verb|remove_equivalent()| removes redundant codes that share the same
$\llbracket n,k,d\rrbracket $ parameters as per Eq.~(\ref{eqiv}). The
\verb|rank()| function computes the rank of matrices over
$\mathbb{F}_2$, and the \verb|is_connected()| function checks if the
code's Tanner graph is connected. Details of the connectivity test can
be found in~\cite{Bravyi_2024}. We focus on connected Tanner graphs
because codes with disconnected Tanner graphs typically have lower
distances.  This \verb|distance_upperbound()| function estimates the
code distance using a threshold $\tau_d$. In essence, the distance can
be estimated using any decoding algorithm by applying errors to a code
word and checking the decoding result to see if it is a logical
error. We conduct multiple trials and track the lowest-weight error
that results in a logical error.  This gives an upper bound of
$d$. Here we used a BP-OSD decoder with 1,000 trials. The threshold
$\tau_d$ is used to skip the rest of trials whenever an error with
weight lower than $\tau_d$ is found, saving computation time. We
employ a similar approach by introducing a threshold $\tau_k$. Since
computing the upper bound of $d$ is considerably slower, we first
calculate $k$. If $k<\tau_k$, the code is discarded without proceeding
to the more time-consuming distance calculation. After the search
process, we can select codes of interest to calculate their exact
distance using an integer programming solver.  As mentioned above, BB
codes have symmetric distances for $X$ and $Z$ errors.  Therefore, it
is sufficient to find the $X$ distance to determine the distance of
the code. It is important to note that, although our search focused on
codes with weight-6 stabilizers of the form specified in
Eq.(\ref{ibm_form}), the algorithm can be adapted to search for codes
with different weights or forms by modifying its input parameters.

A selection of codes found by Algorithm~\ref{alg:bb} is shown in
Table~\ref{bbcodes}. Notably, Eberhardt et
al.~\cite{eberhardt2024logical} propose a
$\llbracket 108, 16, 6\rrbracket $ code and a
$\llbracket 162, 24, 6\rrbracket $ code, which are precisely two and
three times the parameters $n, k$ of our
$\llbracket 54, 8, 6\rrbracket $ BB code, respectively. The
relationship among these three codes remain an area for future
research.
\begin{algorithm}[hbt!]
\caption{Algorithm to search for BB codes}\label{alg:bb}
\SetKwComment{Comment}{/* }{ */}
\SetKwInOut{Input}{Input}
\Input{$l, m, \tau_k, \tau_d$}
\KwResult{codes of parameters $\llbracket 2lm,\geq \tau_k, \leq\hat{d}\rrbracket $}
Generate all polynomial pairs of form Eq.~(\ref{ibm_form}) $L \gets [(a_1(x,y), b_1(x,y)), ...]$\;
$L' \gets \FuncSty{remove\_equivalent}(L)$ \Comment*[r]{Remove codes with the same parameters}
\For{$i\gets 1$ \KwTo $|L'|$}{
    \uIf{\FuncSty{is\_connected}$(a_i(x,y), b_i(x,y)))$}{
        $H_X, H_Z$ = \FuncSty{BB\_matrices}$(a_i(x,y), b_i(x,y)))$\;
        $k\gets 2lm-2\FuncSty{rank}(H_X)$\;
        \uIf{$k<\tau_k$}{
            \KwSty{continue} \;               
        }
        \Else{
            $\hat{d} \gets \FuncSty{distance\_upperbound}(H_X, H_Z, \tau_d)$\;
        }
  }
  \Else{
    \KwSty{continue} \;
  }
    }
\end{algorithm}

\begin{table}[H]
\centering
\caption{Some Novel codes found by Algorithm 1}
\label{bbcodes}
\begin{tabular}{|c|c|>{\centering\arraybackslash}m{2cm}|c|c|}
    \hline
    $l$&$m$ & $a(x,y)$ $b(x,y)$ & $\llbracket n,k,d\rrbracket $ & $kd^2/n$ \\\hline
    3&9 & $1+y^2+y^4$ $y^3+x^1+x^2$ &  $\llbracket 54, 8, 6\rrbracket $ &5.33\\ \hline 
    7&7 & $x^3+y^5+y^6$ $y^2+x^3+x^5$ &  $\llbracket 98, 6, 12\rrbracket $ &8.82\\ \hline 
    3&21 & $1+y^2+y^{10}$ $y^3+x+x^2$ &  $\llbracket 126, 8, 10\rrbracket $ &6.35\\ \hline 
    5&15 & $1+y^6+y^8$ $y^5+x+x^4$ &  $\llbracket 150, 16, 8\rrbracket $& 6.83 \\ \hline 
    3&27 & $1+y^{10}+y^{14}$ $y^{12}+x+x^{2}$ & $\llbracket 162, 8, 14\rrbracket $ &9.68 \\ \hline 
    6&15 & $x^3+y+y^2$ $y^6+x^4+x^{5}$ & $\llbracket 180, 8, 16\rrbracket $ &11.38\\ \hline 
\end{tabular}
\end{table}

\subsection{Coprime Construction}
Based on the commutativity of matrices $x$ and $y$, one can construct
valid CSS codes using various polynomial forms other than
Eq.~(\ref{ibm_form}). For example, we can construct more BB codes by
allowing mixed terms or different numbers of pure $x$- and $y$-
terms. However, this generalization significantly increases the search
space. Especially when using the polynomial form in
Eq.~(\ref{ibm_form}), finding codes with desirable $k$ and $d$ is
already computationally expensive. Thus, we propose a different
construction that not only yields new codes and reduces the search
space but also guarantees to find codes with a predetermined $k$.

We diverge slightly from the original BB codes by letting $l, m$ be
two coprime numbers and $\pi = xy$ to define the polynomials of
coprime-BB codes as follows:
\begin{equation}
    \begin{split}
    &a(\pi)=\sum a_i \pi^i, \quad b(\pi)=\sum b_j \pi^j, \\
    &i,j\in\{0,1,\dots, lm-1\}, a_i, b_j\in\{0,1\}.
    \label{cbb_form}
\end{split}
\end{equation}

{\em Dimension of coprime-BB codes ---} It is easy to verify that
$\langle xy \rangle$ is a cyclic group of order $lm$, thus any
monomial in $\{x^iy^j\:|\: 0\leq i< l,\: 0\leq j<m \}$ can be
expressed as a power of $xy$. Therefore, any polynomials that define
the BB code with coprime $l$ and $m$ can be expressed by univariate
polynomials $a(\pi)$ and $b(\pi)$ and thus in the form of
Eq.~(\ref{cbb_form}). Let
$g(\pi)=\text{GCD}(a(\pi), b(\pi), \pi^{lm}+1)$, where GCD is the
greatest common divisor. Let $\deg g(\pi)$ be the degree of polynomial
$g(\pi)$. The BB code defined by $a(\pi)$ and $b(\pi)$ then has
dimension
\begin{equation}
    k=2\deg g(\pi). 
\end{equation}

\textbf{Proof}: As mentioned above, when $l$ and $m$ are two coprime
integers, any polynomial in $\mathbb{F}_2[x,y]/(x^l+1, y^m+1)$ can be
expressed in $\mathbb{F}_2[\pi]/(\pi^{lm}+1)$. We interpret each
column of the parity-check matrix as a polynomial by taking the column
entries as monomial coefficients. The rest of the proof is similar to
Proposition 1 in~\cite{panteleev2021degenerate}. Given the column
space of $H_X$ is equal to
\begin{equation}
\begin{split}
        \text{colsp}(H_X)&= \{H_X\bm{x}|\bm{x}\in \mathbb{F}_2^{2lm}\} \\
        &= \{A\bm{u}+B\bm{v} | \bm{u},\bm{v}\in \mathbb{F}_2^{lm}\},
\end{split}
\end{equation}
it can be represented in terms of polynomials
\begin{equation}
    \begin{split}
        \text{colsp}(H_X)&=\{a(\pi)u(\pi)+b(\pi)v(\pi) |\\
        & u(\pi),v(\pi)\in\mathbb{F}_2[\pi]/(\pi^{lm}+1)\}.
    \end{split}
\end{equation}
Since $R=\mathbb{F}_2[\pi]/(\pi^{lm}+1)$ is a univariate polynomial
ring, $a(\pi)R$ and $b(\pi)R$ are principal ideals. Thus,
$\text{colsp}(H_X) $ is an principal ideal and is generated by
$g(\pi)$ and $\text{rank}(H_X)=\dim \text{colsp}(H_X)=lm-\deg
g(\pi)$. Therefore, the dimension is given by
\begin{equation}
\begin{split}
\label{dim_cbb}
        k&=2lm - 2\text{rank}(H_X)\\
    &=2lm-2(lm-\deg g(\pi))=2\deg g(\pi).
\end{split}
\end{equation}\qed

The code equivalence described above for BB codes also holds for
coprime-BB codes. However, we want to add another rule for coprime-BB
codes that is easy to implement and can further reduce the search
space by $1/(lm)^2$.

{\em Equivalence for coprime-BB code ---} Let $\mathcal{C}$ be the
code defined by polynomials $a(\pi)$ and $b(\pi)$. The code
$\mathcal{C}'$ defined by the polynomials $\pi^i a(\pi)$ and
$\pi^j b(\pi)$ has the same parameters as $\mathcal{C}$.

{\bf Proof:} From Eq.~(\ref{dim_cbb}), we know that $\mathcal{C}$ and
$\mathcal{C}'$ have the same dimension $k$ because $\pi^i$ and $\pi^j$
are not factors of $\pi^{lm}+1$, thus multiplying the polynomials by
$\pi^i$ and $\pi^j$ does not affect their greatest common divisor.
Assume $\bm{p}=(\bm{\alpha}|\bm{\beta})$ is the binary vector form of
a logical $Z$ operator of $\mathcal{C}$, and let
$A=a(\pi), B = b(\pi), A' = \pi^ia(\pi), B'=\pi^j b(\pi)$ be the
matrices corresponding the polynomials. By definition, a logical $Z$
operator satisfies
\begin{equation}
    A\alpha + B\beta = \bm{0}.
\end{equation}

Now consider a binary vector
$\bm{p}'=(\pi^{-i}\bm{\alpha}|\pi^{-j}\bm{\beta})$. Since $\pi$ is a
permutation matrix (and, hence, invertible), multiplying by $\pi$ does
not change the Hamming weight. Therefore, $\bm{p}'$ has the same
weight as $\bm{p}$. Moreover, all matrices here commute by
construction, implying
\begin{equation}
\begin{split}
        A'\pi^{-i}\bm{\alpha}+B'\pi^{-j}\bm{\beta}=&A\pi^{i}\pi^{-i}\bm{\alpha}+B\pi^{j}\pi^{-j}\bm{\beta}\\ =& A\alpha + B\beta =\bm{0}.
\end{split}
\end{equation}
Thus, $\bm{p}'$ is a logical $Z$ operator of
$\mathcal{C}'$. Similarly, we can prove this property for a logical
$X$ operator by conversely implying that $\mathcal{C}'$ and
$\mathcal{C}$ are equivalent in terms of their parameters. \qed
\begin{table*}[!ht]
\centering
\caption{Some novel coprime-BB codes found by Algorithm 2.}
\label{ppbb}
\begin{tabular}{|c|c|c|c|c|c|}
    \hline
    $l$ & $m$ & $a(\pi), a(x,y)$ & $b(\pi), b(x,y)$ & $\llbracket n,k,d\rrbracket $ &$kd^2/n$\\\hline
    3 & 5 & $1+\pi+\pi^2=1+xy+(xy)^2$  & $1+\pi^2+\pi^{7}=1+(xy)^2+xy^2$ &  $\llbracket 30, 4, 6\rrbracket $ &4.8 \\ \hline 
    3 & 7 & $1+\pi^2+\pi^3=1+(xy)^2+y^3 $ & $1+\pi^2+\pi^{10}=1+(xy)^2+xy^3$ &  $\llbracket 42, 6, 6\rrbracket $ &5.14 \\ \hline 
    5 & 7 & $1+\pi+\pi^5=1+(xy)+y^5$  & $1+\pi+\pi^{12}=1+xy+x^2y^5$ &  $\llbracket 70, 6, 8\rrbracket $&5.49 \\ \hline 
    2 & 27 & $1+\pi^{3}+\pi^{42}=1+xy^3+y^{15}$ & $1+\pi^{6}+\pi^{39}=1+y^{6}+xy^{12}$ & $\llbracket 108, 12, 6\rrbracket $&4 \\ \hline
    7 & 9 & $1+\pi+\pi^{58}=1+xy+x^2y^4$ & $1+\pi^{13}+\pi^{41}=1+x^6y^4+x^6y^5$ & $\llbracket 126, 12, 10\rrbracket $&9.52 \\ \hline
    7 & 11 & $1+\pi+\pi^{31}=1+xy+x^3y^9$ & $1+\pi^{19}+\pi^{53}=1+x^5y^8+x^4y^9$ & $\llbracket 154, 6, 16\rrbracket $&9.97\\ \hline

\end{tabular}
\vspace*{-\baselineskip}
\end{table*}

\begin{algorithm}[hbt!]
\caption{An algorithm to search for BB codes with the new form of polynomials.}
\label{alg:ppbb}
\SetKwInOut{Input}{Input}
\Input{$l, m, \tau_d, \tau_k$ } 
\KwResult{codes of parameters $\llbracket 2lm,\geq \tau_k,\leq \hat{d} \rrbracket$}
$G\gets \FuncSty{factors}(\pi^{lm}+1)$\; \Comment {Find all factors over $\mathbb{F}_2$}
\For{$g(\pi) \in G$}{
    \uIf{$2\deg g(\pi)<\tau_k$}{\KwSty{continue}\;}
    $C\gets$ all polynomials $f(\pi)$ in $\mathbb{F}_2[\pi]/(\pi^{lm}+1)$ s.t. $\text{wt}(f(\pi))=3$\;
\Comment {Or > 3 to get higher weight codes}
$C' \gets$ all polynomials $c(\pi)$ in $C$ s.t.  $c(\pi)\mod g(\pi) = 0$\;
$L \gets$ all combinations $(a(\pi), b(\pi))$ chosen from $C'$ s.t. $\text{GCD}(a(\pi), b(\pi)) = g(\pi)$\;
$L' \gets \FuncSty{remove\_equivalent}(L)$\;
\For{$i\gets 1$ \KwTo $|L'|$}{
        $H_X, H_Z$ = \FuncSty{BB\_matrices}$(a_i(\pi), b_i(\pi)))$\;
        $k\gets 2\deg g(\pi)$\;
        $\hat{d} \gets \FuncSty{distance\_upperbound}(H_X, H_Z, \tau_d)$\;
  \Else{
    \KwSty{continue} \;
  }
}

    }
\end{algorithm}

Coprime-BB codes can also be viewed as a special case of generalized
bicycle (GB) codes~\cite{panteleev2021degenerate}. In GB codes,
polynomials are identified as sums of cyclic shift matrices. In
contrast, coprime-BB codes identify polynomials as sums of the
Kronecker products of two cyclic shift matrices with coprime
dimensions. By adapting the new construction to
Algorithm~\ref{alg:bb}, we propose Algorithm~\ref{alg:ppbb}.
The latter algorithm results in a significantly reduced search space
as only coprimes
and codes with desired $k$ are being considered.
Using Algorithm~\ref{alg:ppbb}, we found a number of interesting
coprime-BB codes, which are shown in Table~\ref{ppbb} with their
polynomials in the bivariate form. We also noted that
in~\cite{panteleev2021degenerate}, the author found a
$\llbracket 126, 12, 10\rrbracket$ with different polynomials, i.e.,
we are not the first to discover this particular code, yet our search
algorithm identified it based on a different basis.  For large $l,m$,
the results are not exhaustive as numerous polynomials meet the
condition $\text{GCD}(a(\pi), b(\pi)) = p(\pi)$. As the algorithm
outputs codes with the same $k$, the results shown were obtained by
running the search algorithm for a short duration and selecting the
output code with the highest estimated $d$.

\begin{figure}[!t]
    \centering
    \includegraphics[width=3.2 in]{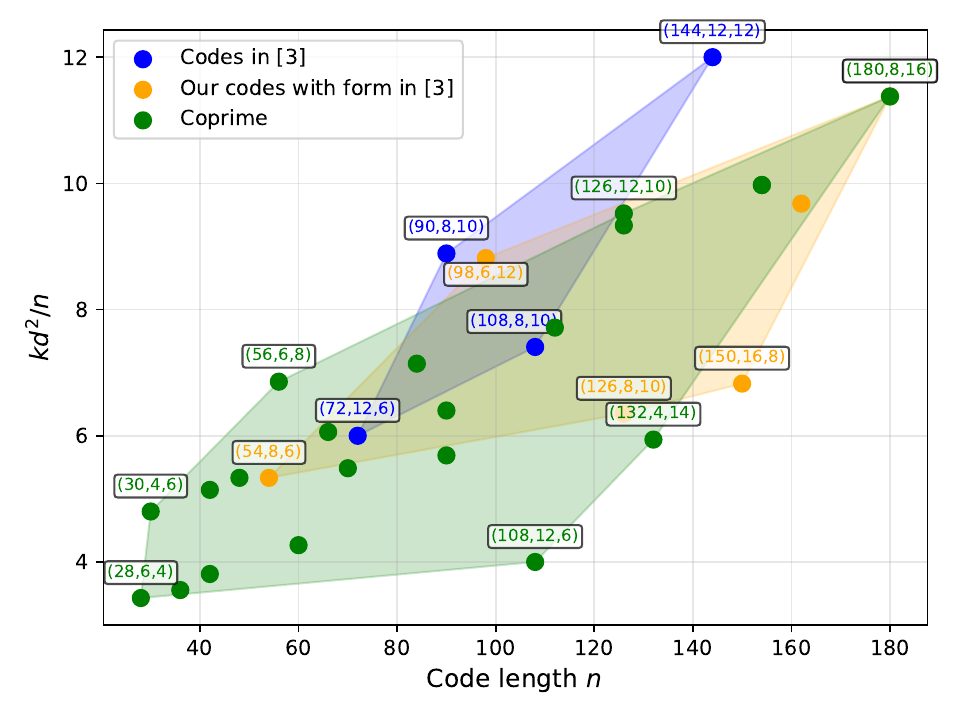}
    \caption{Comparison of quantum codes using the normalized metric
      $kd^2/n$ versus code length $n$ for $n < 200$. The plot includes
      codes from~\cite{Bravyi_2024} (``Codes in~\cite{Bravyi_2024}''),
      our constructions matching the form in~\cite{Bravyi_2024}, and
      newly discovered coprime-based codes.}
    \label{nkdplot}
\vspace*{-\baselineskip}
\end{figure}

Additionally, we visualize our newly discovered codes alongside those
proposed in~\cite{Bravyi_2024} in Fig.~\ref{nkdplot} (restricted to
$n < 200$). The codes are plotted using the metric $kd^2/n$ against
the code length $n$, which highlights improvements over surface codes,
characterized by $n \propto kd^2$. As shown in the figure, our new
constructions expand the spectrum of available codes at short lengths,
offering greater flexibility in code selection suitable for near-term
module sizes suitable for distributed quantum
computing.

We observe that the $\llbracket 126, 12, 10\rrbracket $ code offers
the highest $d$ and one of the highest $k$, which is comparable to the
$\llbracket 144,12,12\rrbracket $ ``gross'' code proposed
in~\cite{Bravyi_2024}, making it suitable for scenarios where error
rates are moderate to high, and ample qubit resources exist.
The $\llbracket 42, 6, 6\rrbracket $ code has a low $d$ but provides
the highest rate of these codes and is best suited for scenarios with
low error rates or where hardware resources are limited. The
$\llbracket 70, 6, 8\rrbracket $ code offers a balanced trade-off
between error correction capabilities and code length, making it a
versatile option for environments with moderate physical error rates
and resource constraints.

\section{A Novel Layout Optimization for Coprime-BB Codes}
\begin{figure}[!t]
    \centering
    \includegraphics[width=3.3 in]{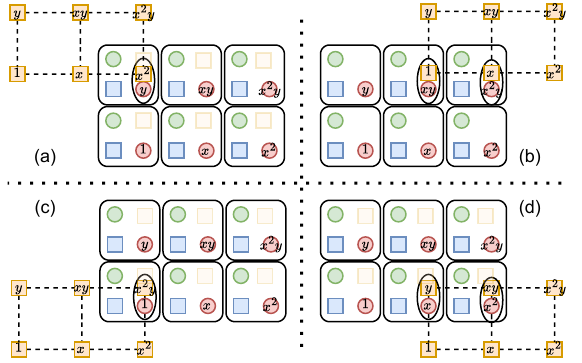}
    \caption{Movements required to perform an \( xy \) check (\( l =
      3, m = 2 \)) on the BB layout from~\cite{viszlai2023matching}.
      The required CNOTs are between the following pairs: 
      \( (X_1, L_{xy}) \), 
      \( (X_y, L_x) \), 
      \( (X_x, L_{x^2y}) \), 
      \( (X_{xy}, L_{x^2}) \), 
      \( (X_{x^2}, L_y) \), 
      and \( (X_{x^2y}, L_1) \).
      Four movement steps are needed, as two periodic boundary
      conditions are crossed during the process.}
    \label{4mov}
\end{figure}

As a subclass of BB codes, coprime-BB codes distinguish themselves by
one major feature: Almost every monomial in the polynomials that
define them is ``mixed'', i.e., each monomial contains both $x$ and
$y$. In the BB layout by~\cite{viszlai2023matching}, we need four
steps to perform the CNOTs required by a mixed monomial as shown in
Fig.~\ref{4mov}. This is because the mixed term check introduces two
types of periodicity, one vertically and one horizontally, rather than
just one type of periodicity for codes defined by
Eq.~(\ref{ibm_form}). However, a {\em global} Rydberg laser pulse
introduces errors to {\em all} atoms after each step of movement, not
just those within the blockade
radius~\cite{bochenqubit}. Consequently, codes with mixed terms are
more prone to error, since checking them requires additional global
laser pulses and thereby increasing the likelihood of errors on idle
qubits.

\begin{figure}[!t]
    
    \centering
    \includegraphics[width=2.7 in]{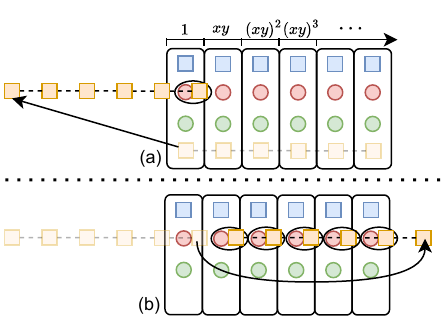}
    \caption{The movements to perform a $xy$ check using the proposed
      CBB layout ($l=3, m=2$). }
    \label{cbb_layout}
\end{figure}

Despite this drawback, coprime-BB codes have another useful property:
Every monomial is of the form of $(xy)^i$. This structure naturally
suggests a one-dimensional arrangement of atoms, which can avoid
multiple movements and thereby help reduce the error rates. Therefore,
we propose a {\em novel} layout design for coprime-BB codes using this
property. As illustrated in Fig.~\ref{cbb_layout}, we organize the
$X, Z, L$ and $R$ qubits with the same subscript vertically, and we
place them horizontally in the order of subscripts
$1, xy, (xy)^2, ..., (xy)^{lm-1}$. By doing so, every move of form
$(xy)^i$ can be seen as a cyclic shift horizontally, i.e., it can be
performed in two moves: Move left by $lm-i$ subgrids and then move
right by $lm$ subgrids. We call this novel layout the {\em ``CBB
  layout''} for the remainder of this paper. Since $l$ and $m$ are
coprime integers, we can ensure that this layout covers all qubits
without repetition in $lm$ columns. As a result, any syndrome
extraction can be accomplished by moving the ancilla qubits
horizontally (if we ignore the minor vertical movement between the
atoms with the same subscript).

\begin{figure}[!t]
    \centering
    \includegraphics[width=3.3 in]{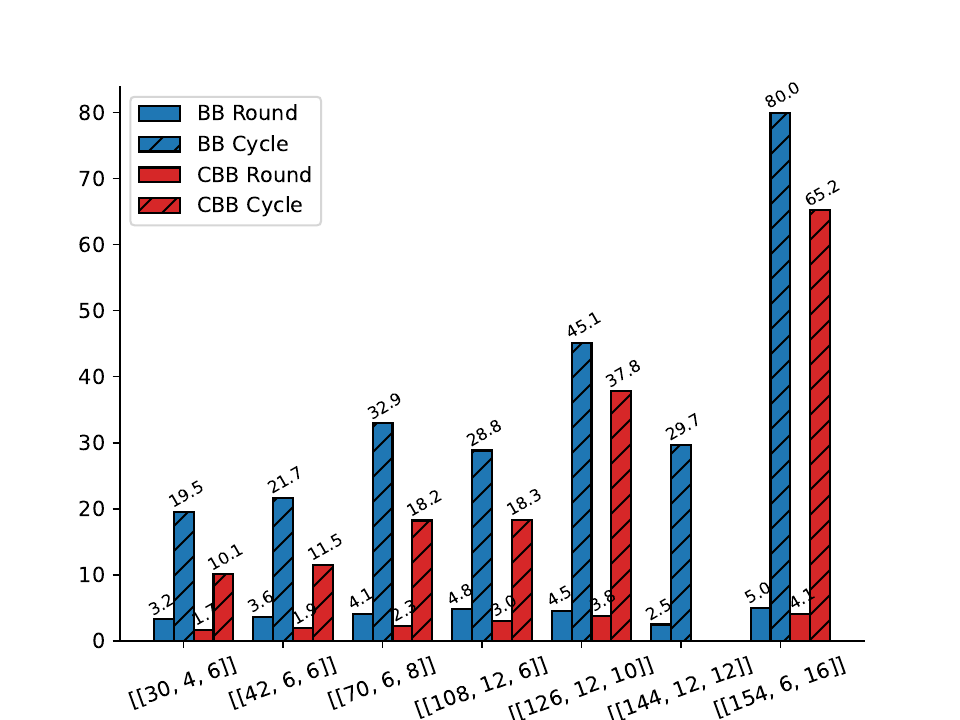}
    \caption{Movement time of different codes and layouts. The CBB layout for $\llbracket144,12,12\rrbracket$ code is omitted because the CBB layout only applies for coprime-BB codes.}
    \label{move_time}
\end{figure}

It is also worth noting that the move distance to perform a mixed-term
check in the BB layout is at least $4(l+m)$, assuming that the atom
distance is 1. In contrast, the move distance of the CBB layout is at
least $2lm$. Notice that the latter term starts lower but grows faster
than the former as the code length increases, but the move-time
cross-over point is beyond the scale of studied code sizes today,
i.e., in practice CBB layouts perform better for realistic
codes). Figure~\ref{move_time} compares the corresponding movement
times for coprime-BB codes of different layouts, where a cycle refers
to the time needed for one syndrome-extraction cycle, and a round
consists of $d$ such cycles. We also add the movement time of the
$\llbracket 144,12,12\rrbracket$ code under scheduling proposed
in~\cite{viszlai2023matching} as a reference. Following the approach
in~\cite{xu2024constant, viszlai2023matching}, we calculate movement
times assuming an atom spacing of $5\mu m$ and an acceleration of
$0.02 \mu m/\mu s^2$. The time for a $(\Delta x, \Delta y)$ move,
which proceeds along along Manhattan path, is given by
\begin{equation}
    \sqrt{6\Delta x/0.02} + \sqrt{6\Delta y/0.02}.
    \label{move_eq}
\end{equation}
The optimal route to complete all monomial checks is computed using a
traveling salesman problem (TSP) solver on the Manhattan distance
metric. From Figure~\ref{move_time} we observe that the proposed CBB
layout has a lower cycle time despite having a higher move
distance. This is because the BB layout requires many short moves,
whereas the CBB layout relies on fewer but longer moves, which is more
efficient when factoring in acceleration overhead, i.e., AOD movement
first accelerates to the mid-point before it slows down as suggested
in Eq.~(\ref{move_eq}). This benefit of CBB is becoming smaller as the
code length becomes large, but we have not observed this phenomenon in
the codes tested here.

\textbf{Discussion: }In the CBB layout, qubits are
  arranged into long stripes forming a $1\times lm$ grid, rather than
  a $l\times m$ rectangular array. This geometry places increased
  demands on acousto-optic deflectors, since the larger number of
  columns requires a correspondingly higher number of resolvable
  radio-frequency tones, which may constrain the maximum achievable
  array size due to finite AOD bandwidth and spot-resolution
  limits. However, coprime-BB codes are known to be asymptotically
  limited~\cite{postema2025existence}, and are therefore not intended
  for very large code distances. As a result, our focus is on
  short-to-medium code lengths, for which the required number of sites
  per row remains moderate. Recent experimental demonstrations of
  neutral-atom arrays have achieved grids as large as $120\times12$
  \cite{Chiu2025} suggesting that resolving on the order of $10^2$
  atoms per row is feasible with current technology.

 In addition, ancilla transport in the CBB layout
  typically involves longer distances per move than in BB
  layouts. Although our analysis indicates that the total number of
  movement operations, as well as the overall transport time for the
  codes considered here, is reduced, this reduction is achieved at the
  cost of higher average transport velocities. As discussed in
  \cite{PhysRevApplied.19.054032, Roberts:14}, the start-stop jerking
  effect could be a major source of heating and atom loss. The
  trade-off between fewer start–stop events and increased velocity
  makes the net impact on transport-induced noise, such as motional
  heating, difficult to assess without detailed experimental
  characterization. In summary, the CBB layout is more suitable for
  short-to-medium code lengths on near-term devices.

\section{Numerical Results}
In this section, we evaluate the error rates of the newly found codes
through theoretical analysis and numerical simulation. We begin by
illustrating the performance of the codes under the code-capacity
model using Monte Carlo simulations, followed by a simulation based on
the circuit-based noise model using Stim~\cite{gidney2021stim}. All
simulation results are gathered for 100 or more logical errors,
ensuring that error bars remain below approximately $\pm10\%$.

\subsection{Code-Capacity Model}
\begin{figure}[!t]
    \centering
    \includegraphics[width=3.3 in]{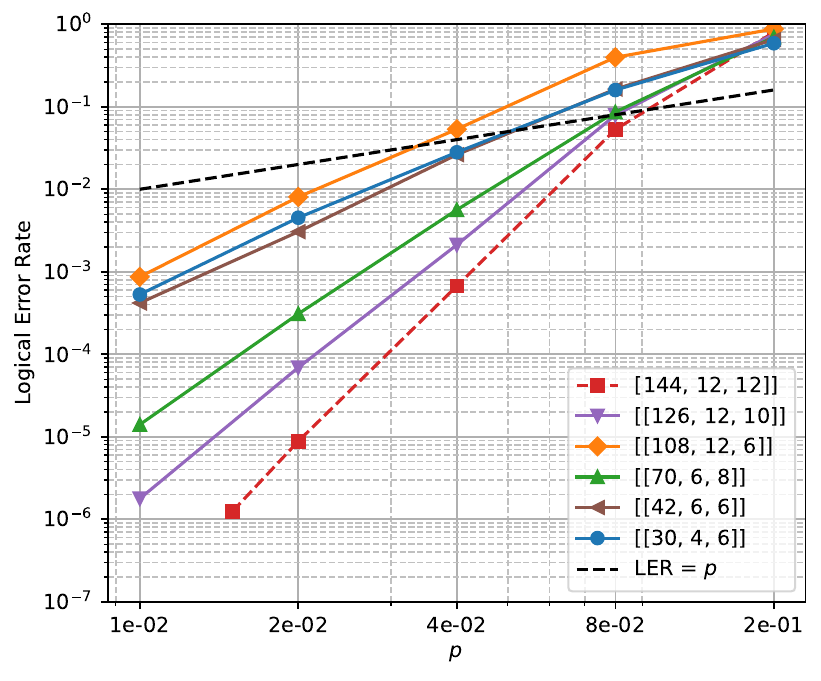}
    \caption{Logical error rates of different codes under code
      capacity model. The lowest error rate point for
      $\llbracket 144, 12, 12\rrbracket$ is simulated under physical error rate
      $p=1.5\times 10^{-2}$.}
    \label{capacity_fer}
\end{figure}

In the code-capacity model, we assume that all gate operations and
measurements are perfect. $X,Y$ and $Z$ errors are applied with $p/3$
independently for each data qubit. In Fig.~\ref{capacity_fer}, the
logical error rates (y-axis) of proposed coprime-BB codes and the
$\llbracket 144,12,12\rrbracket $ BB code are depicted for different
physical error rates (x-axis). The decoder used in the simulation is
BP-OSD~\cite{panteleev2021degenerate, roffe2020} with a maximum of
10,000 min-sum (MS) iterations, variable scaling factor, and the
``OSD\_CS'' method of order 10. As expected, codes with a larger
distance $d$ generally exhibit better performance than those with a
smaller $d$. Among codes with the same distance $d$, such as
$\llbracket 108, 12, 6\rrbracket $, and
$\llbracket 30, 4, 6\rrbracket $, the code with shorter code length
tends to have a lower logical error rate as longer codes have more
sources of errors.

\subsection{Circuit-Level Noise Model On Cold Atoms Arrays}

Under the circuit-level noise model, we consider a more realistic
scenario in which errors can occur during any operation, except for
classical processing. We adopt the common practice as described
in~\cite{viszlai2023matching}, where the syndrome extraction is
performed by multiple rounds to compensate for measurement errors. In
each round, noise is introduced independently on each measurement and
gate operation with a certain probability. The number of rounds is set
equal to the distance of the code. After performing the desired
rounds, the syndrome history is fed into the decoder to estimate the
final error. In our simulations, we assume that the single-qubit gate
error rate ($e_{1g})$, two-qubit gate error rate ($e_{2g})$, and
readout error rate ($e_r$) all share a common probability $p$. We
model single- and two-qubit gate errors using depolarizing channels:
for single-qubit gates, each non-identity Pauli error occurs with
probability $p/3$, while for two-qubit gates, it occurs with
probability $p/15$. Idle errors are applied after each move and are
determined based on the device's relaxation time ($T_1$), dephasing
time ($T_2$), and the time required for each move
(see~\cite{viszlai2023matching} for details).  We will use
$T_1=T_2=1 \:s$ throughout this work. Read-out errors are applied
using $X$ or $Z$ flip with probability $p$ based on the measurement
basis before each measurement.

According to~\cite{bochenqubit}, a global Rydberg laser introduces
noise on all atoms, including those not actively participating in
entanglement operations. To model this effect, we introduce a
coefficient $c$. Each time a two-qubit global gate is applied, in
addition to the two-qubit depolarization noise on the interacting
qubit pairs, we apply an additional single-qubit depolarization noise
with probability $c\cdot e_{2g}$ to all qubits.
\begin{figure}[!t]
    \centering
    \includegraphics[width=3.3 in]{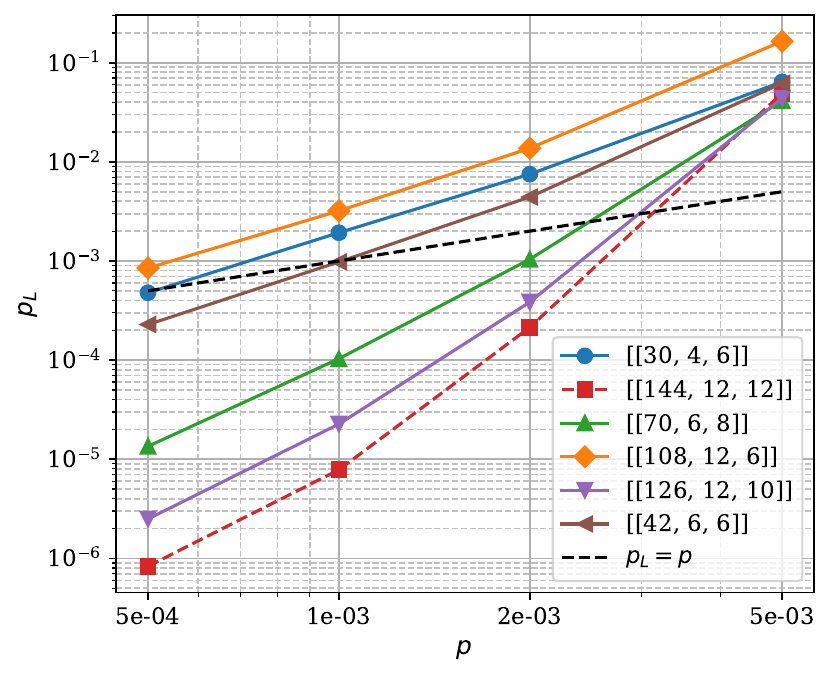}
    \caption{Logical error rates per round ($p_L$) of selected codes under circuit-level noise model with different physical error rate ($p$).}
    \label{circuit_fer}
\end{figure}
\begin{figure*}[!ht]
     \centering
    \begin{subfigure}{0.32\textwidth}
  \includegraphics[width=\textwidth]{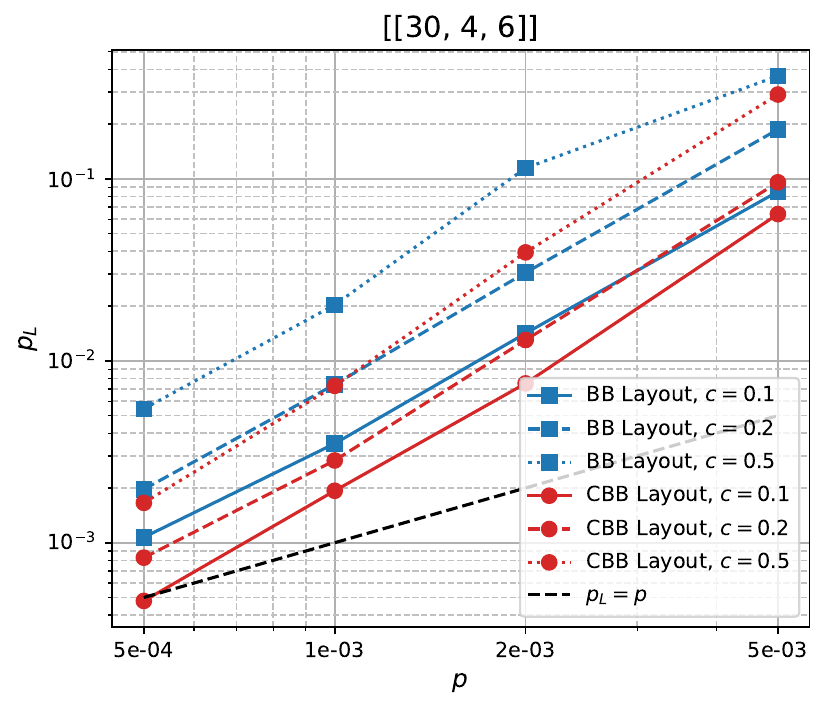}
  \caption{}
  \end{subfigure}
  \hfill
  \begin{subfigure}{0.32\textwidth}
  \includegraphics[width=\textwidth]{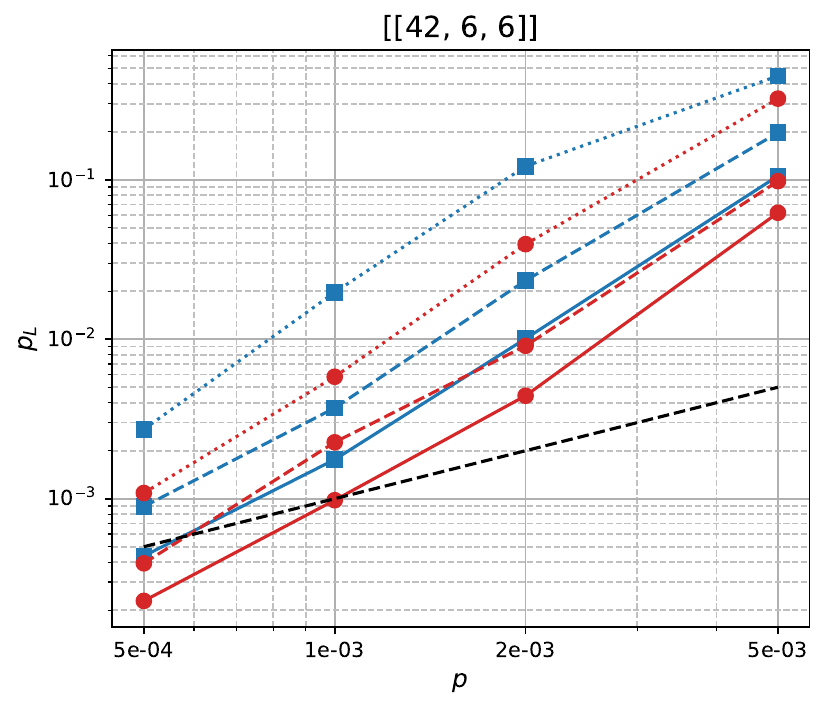}
    \caption{}

  \end{subfigure} 
  \hfill
  \begin{subfigure}{0.32\textwidth}
  \includegraphics[width=\textwidth]{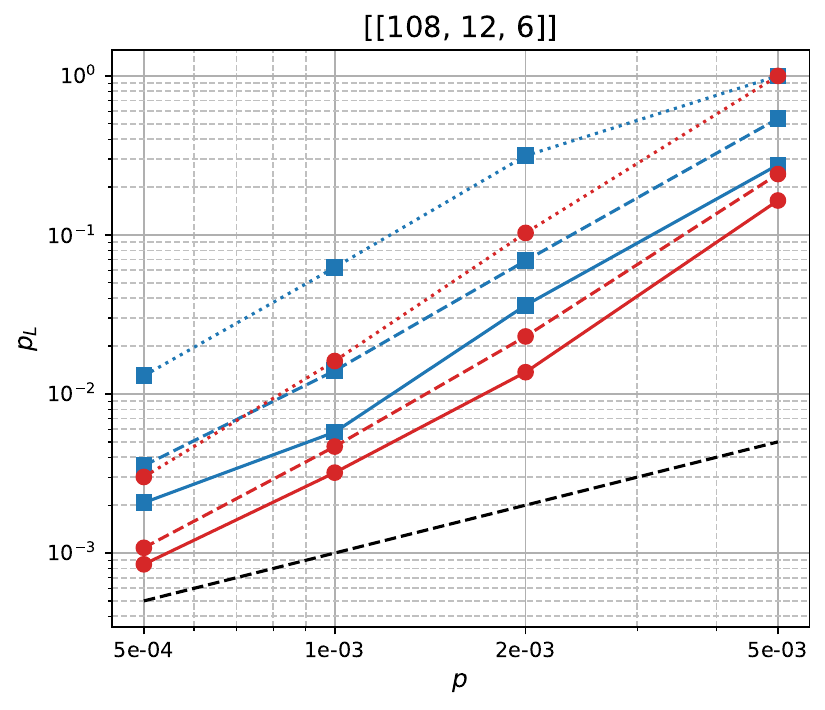}
  \caption{}
  \end{subfigure}
  \\
    \begin{subfigure}{0.32\textwidth}
  \includegraphics[width=\textwidth]{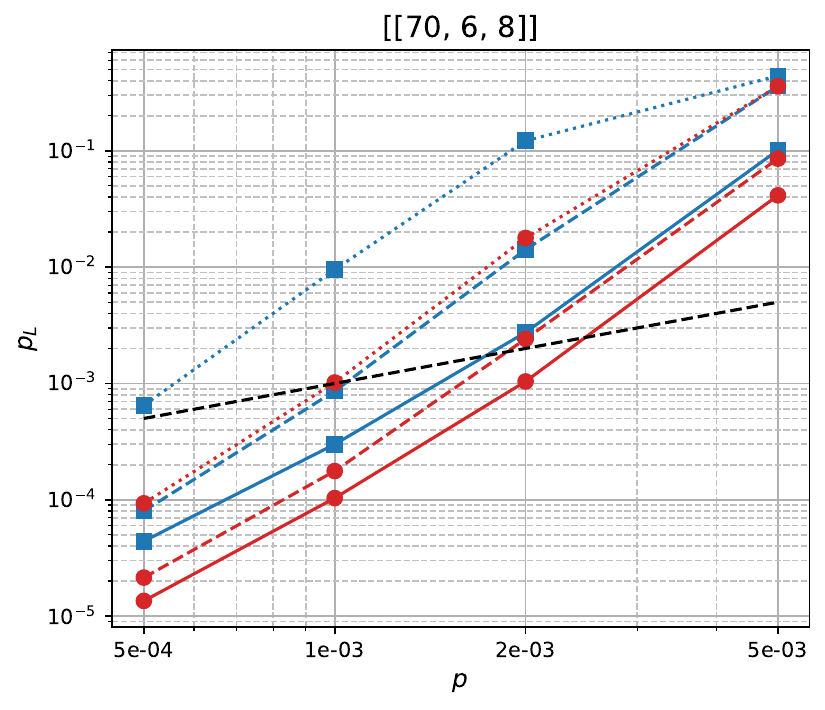}
  \caption{}
  \end{subfigure}
  \hfill
  \begin{subfigure}{0.32\textwidth}
  \includegraphics[width=\textwidth]{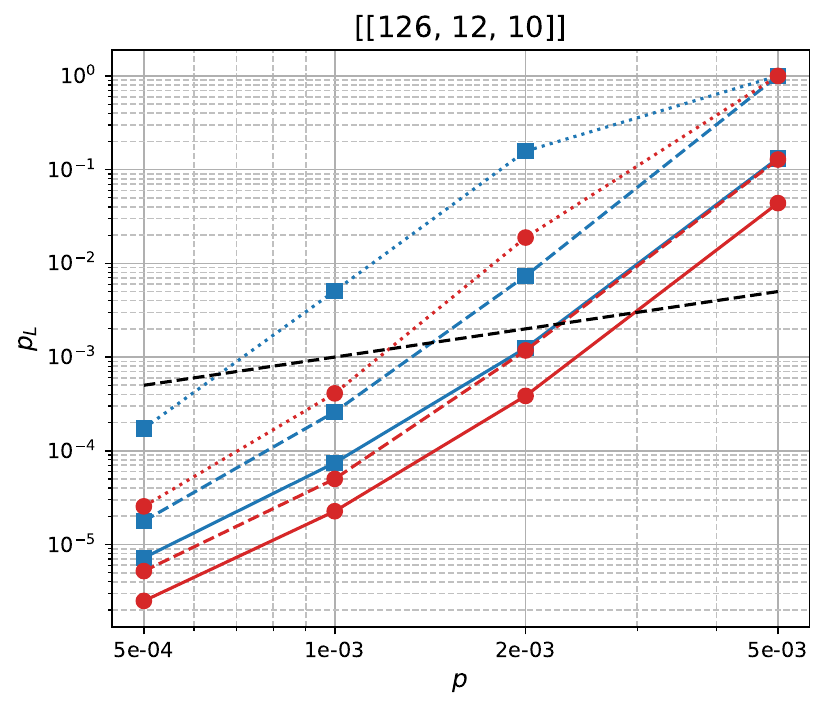}
  \caption{}
  \end{subfigure} 
  \hfill
  \begin{subfigure}{0.32\textwidth}
  \includegraphics[width=\textwidth]{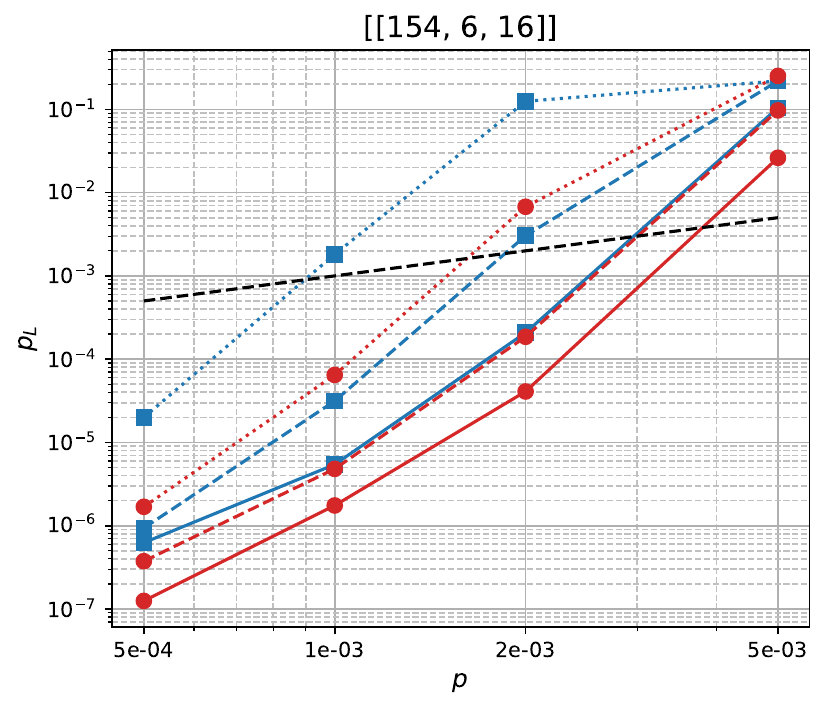}
  \caption{}
  \end{subfigure}
  \caption{The circuit level simulations for different coprime-BB
    codes under different coefficients $c$ and using different
    layouts. $p_L$ is logical error rate per round and $p$ is physical error rate. (a-f) share the same legend as in (a). (a-c) show the
    comparison for lower distance coprime-BB codes, and (d-f) show
    that for codes with higher distances.}
     \label{c_compare}
\end{figure*}
Fig.~\ref{circuit_fer} presents the logical error rate for selected
coprime-BB codes across a range of physical qubit error rates. Except
for the $\llbracket 144, 12, 12\rrbracket$ BB code, which is
incompatible with the CBB layout, all codes are arranged in the CBB
layout. A global layer error rate coefficient of $c=0.1$ is used for
all codes. Unlike the code-capacity model, we evaluate performance via
the logical error rate per cycle as
\begin{equation}
    p_L=1-(1-\frac{N_e}{N})^{1/d},
\end{equation}
where $N_e$ is the number of logical errors and $N$ is the number of
simulations. 
%

Similar to the code-capacity model, codes with higher $d$ tend to
achieve a lower logical error rate per cycle. However, we observe that
the $\llbracket 42, 6, 6\rrbracket $ code exhibits a lower logical
error rate than $\llbracket 30, 4, 6\rrbracket $, which has a shorter
code length and the same $d$. This discrepancy arises because they
have different circuit distances $d_{circ}$, defined as the minimum
number of error mechanisms to flip a logical observable without
triggering any detection event. Since errors can propagate through the
circuit and effectively increase the number of errors, we usually have
$d_{circ}<d$. Specifically, $\llbracket 30, 4, 6\rrbracket $ has
$d_{circ}=3$, whereas $\llbracket 42, 6, 6\rrbracket $ has
$d_{circ}=4$. In Fig.~\ref{circuit_fer}, we also observe that the
performance gap between codes $\llbracket 144, 12, 12\rrbracket $ and
$\llbracket 126, 12, 10\rrbracket $ is much smaller than in
Fig.~\ref{capacity_fer}. This reduced gap is likely due to the same
reason, i.e., simulations indicate that the
$\llbracket 144, 12, 12\rrbracket $ has a circuit-level distance of
$d_{circ} \leq 10$, while the $\llbracket 126, 12, 10\rrbracket $ has
$d_{circ} \leq 9$.

Figure~\ref{c_compare} presents the logical error rates $p_L$ for
different coprime-BB codes under both the BB and CBB layouts for
global laser error coefficients $c\in\{0.1,0.2,0.5\}$. Across all
tested codes, the proposed CBB layout consistently achieves lower
logical error rates than the BB layout, owing primarily to its shorter
round/cycle time and fewer ancilla movements. These advantages reduce
the number of required global CNOT operations, thereby lowering the
accumulated noise level on idle qubits. The improvement is more
significant for high distance codes, especially when the global noise
coefficient $c$ is high. For instance, we have an error rate reduction
around $1/10$ on the $\llbracket 126, 12, 10\rrbracket$ and
$\llbracket 154, 6, 16\rrbracket$ codes for $c=0.5$. In contrast, at
$c=0.1$, the improvement is $1/2$ and $1/6$, respectively. This is due
to a higher $c$ indicating that each global laser pulse introduces more
noise on all qubits. As the CBB layout saves two global laser
operations per monomial in a polynomial, we can expect even greater
gains for codes with higher polynomial weight.

\section{Related Work}
\label{sec:related}
Recently, various constructions based on qLDPC codes have been
proposed to achieve different objectives. Koukoulekidis et
al.~\cite{koukoulekidis2024smallquantumcodesalgebraic} proposed
algebraic extensions to expand a small GB code into a family of larger
GB codes by selecting a sequence of expansion factors. The authors
also introduced scalable codes that embed the original short codes
into extended codes, enhancing scalability in superconducting
architectures. Voss et al.~\cite{voss2024multivariate} expanded the
concept of BB codes by introducing an additional type of indeterminate
variable, leading to the creation of trivariate bicycle (TB)
codes. These new codes reduce the weight of stabilizers from 6 to 4-5,
making them more practical for hardware implementation. However, it is
important to note that some of these codes exhibit a lower rate or
distance compared to BB codes of similar length. Eberhardt et
al.~\cite{eberhardt2024logical} investigated the algebraic structure
of BB codes and uncovered certain symmetry properties. These
properties allow for the explicit construction of logical operators
and certain fault-tolerant gates for BB codes.  Shaw et
al.~\cite{shaw2024lowering} proposed a ``morphing'' circuit design for
syndrome extraction on BB codes. The proposed circuit design has only
six rounds of CNOT gates instead of seven~\cite{Bravyi_2024}. By
applying the proposed circuit, the authors discovered a new family of
BB codes, including codes that have the same
$\llbracket n,k,d\rrbracket $ parameters as~\cite{Bravyi_2024}. The
authors also provide a sufficient condition for the circuit to be
applied to the other two-block group algebra (2BGA) \cite{PhysRevA.109.022407} codes.

\section{Conclusion and Future Work}
\label{sec:conc}

We developed fast numerical search algorithms to discover BB codes and
introduced a novel construction method using factor polynomials of
$\mathbb{F}_2[\pi]/(\pi^{lm}+1)$, where $l$ and $m$ are coprime
integers. The new construction enables us to know the rate of BB codes
before constructing them. We also introduced a new error model that
accounts for noise of global Rydberg laser pulses affecting
non-interacting qubits and demonstrated the error rates of our newly
discovered codes in simulation. Moreover, we devise a novel and more
efficient mapping of coprime-BB codes tailored to cold atom-array
architectures. Our approach achieves shorter movement times, fewer
moves, and lower error rates on the codes we tested.
These features make the coprime-BB code
a candidate for quantum memory. However, further research is needed to
explore other properties of these codes, such as logical gate
constructions and implementation on superconducting architectures.

\section*{Acknowledgment}
The authors would like to thank Joshua Viszlai, Hanrui Wang, and John
Stack for fruitful discussions on cold atom arrays and error
correction. They also thank Yu-An Chen for identifying errors in the
manuscript.  This work was supported in part by NSF awards
OSI-2410675, PHY-1818914, PHY-2325080, OMA-2120757, CISE-2217020, and
CISE-2316201 as well as DOE DE-SC0025384.

\bibliographystyle{quantum}

\bibliography{main}

\appendix
\section{Other BB Codes Found}
In addition to the BB codes provided so far, we present additional
coprime-BB codes found by Algorithm 1.  Any ``obviously'' inferior
codes found are not included in this table, e.g., codes with the same
$n$ but lower $k$ or $d$.

\begin{table}[H]
\centering
\caption{Selected weight-6 codes found by Algorithm 1.}
\label{bbcodes_app}
{\small\begin{tabular}{|c|c|c|c|c|}
    \hline
    $l$ & $m$ & $a(x,y)$  $b(x,y)$ & $\llbracket n,k,d\rrbracket $ &$kd^2/n$\\\hline
    3 & 3 & $1+x+y$   $1+x^2+y^{2}$ &  $\llbracket 18, 4, 4\rrbracket $ &3.56\\ \hline 
    3 & 6 & $1+y+y^{2}$   $x^3+y+y^2$ &  $\llbracket 36, 8, 4\rrbracket $ &3.56\\ \hline 
    3 & 6 & $x+y^2+y^{3}$   $1+y+x^{2}$ &  $\llbracket 36, 4, 6\rrbracket $&4 \\ \hline 
    3 & 9 & $x+y+y^{3}$   $1 + y^2+x^{2}$ &  $\llbracket 54, 4, 8\rrbracket $&4.74 \\ \hline 
    7 & 14 & $1+y+y^{3}$  $y^7+x+x^{3}$ & $\llbracket 196, 18, 8\rrbracket $&5.87\\ \hline 
    \end{tabular}}
    \end{table}
\section{Other Coprime-BB Codes Found}
\begin{table*}[!h]
\centering
\caption{Selected weight-6 codes found by Algorithm 2}
\label{cbbcodes}
{\small
\begin{tabular}{|c|c|c|c|c|c|}
    \hline
    $l$ & $m$ & $a(\pi), a(a,y)$ & $b(\pi), b(x,y)$ & $\llbracket n,k,d\rrbracket $&$kd^2/n$ \\\hline
    2 & 7 & $1+\pi^1+\pi^3=1+xy+xy^3$  & $1+\pi^1+\pi^{10}=1+xy+y^3$ &  $\llbracket 28, 6, 4\rrbracket $ &3.43\\ \hline 
    2 & 9 & $1+\pi^2+\pi^{10}=1+y^2+y$  & $1+\pi^4+\pi^8=1+y^4+y^8$ &  $\llbracket 36, 8, 4\rrbracket $ &3.56\\ \hline 
    3 & 7 & $1+\pi+\pi^{5}=1+xy+x^2y^5$  & $1+\pi^2+\pi^{10}=1+(xy)^2+xy^3$ &  $\llbracket 42, 10, 4\rrbracket $&3.8 \\ \hline 
    3 & 8 & $1+\pi+\pi^{2}=1+xy+(xy)^2$  & $1+\pi^2+\pi^{10}=1+(xy)^2+xy^2$ &  $\llbracket 48, 4, 8\rrbracket $ &5.33\\ \hline 
    3 & 10 & $1+\pi^2+\pi^{8}=1+(xy)^2+x^2y^8$ & $1+\pi^4+\pi^{16}=1+xy^4+xy^6$ & $\llbracket 60, 16, 4\rrbracket $&4.27 \\ \hline 
    3 & 11 & $1+\pi+\pi^{5}=1+xy+x^2y^5$ & $1+\pi+\pi^{23}=1+xy+x^2y$ & $\llbracket 66, 4, 10\rrbracket $ & 6.06\\ \hline 
    4 & 7 & $1+\pi+\pi^{3}=1+xy+(xy)^3$ & $1+\pi^5+\pi^{11}=1+xy^5+x^3y^4$ & $\llbracket 56, 6, 8\rrbracket $& 6.86\\ \hline 
    5 & 9 & $1+\pi+\pi^{4}=1+xy+(xy)^4$ & $1+\pi^8+\pi^{34}=1+x^3y^8+x^4y^7$ & $\llbracket 90, 4, 12\rrbracket $&6.4\\ \hline 
    5 & 9 & $1+\pi+\pi^{12}=1+xy+x^2y^3$ & $1+\pi^2+\pi^{9}=1+(xy)^2+x^4$ & $\llbracket 90, 8, 8\rrbracket $&5.69\\ \hline 
    6 & 7 & $1+\pi+\pi^{3}=1+xy+(xy)^3$ & $1+\pi^8+\pi^{31}=1+x^2y+xy^3$ & $\llbracket 84, 6, 10\rrbracket $&7.14\\ \hline 
    6 & 11 & $1+\pi+\pi^{2}=1+xy+(xy)^2$ & $1+\pi^{11}+\pi^{28}=1+x^5+x^4y^6$ & $\llbracket 132, 4, 14\rrbracket $&5.94\\ \hline 
    7 & 8 & $1+\pi+\pi^{3}=1+xy+(xy)^3$ & $1+\pi^{5}+\pi^{25}=1+(xy)^5+x^4y$ & $\llbracket 112,6, 12\rrbracket $&7.71\\ \hline
    7 & 9 & $1+\pi^4+\pi^{19}=1+(xy)^4+x^5y$ & $1+\pi^{6}+\pi^{16}=1+(xy)^6+x^2y^7$ & $\llbracket 126,6, 14\rrbracket $&9.33\\ \hline
    9 & 10 & $1+\pi+\pi^{4}=1+xy+(xy)^4$ & $1+\pi^{23}+\pi^{62}=1+x^5y^3+x^8y^2$ & $\llbracket 180, 8, 16\rrbracket $&11.38\\ \hline
    
\end{tabular}}
\end{table*}
\begin{table*}[!h]
\centering
\caption{Selected weight-8 codes found by Algorithm 2}
\label{cbbcodes8}

{\small
\begin{tabular}{|c|c|>{\centering\arraybackslash}m{3.5cm}|>{\centering\arraybackslash}m{3.5cm}|c|c|}
    \hline
    $l$ & $m$ & $a(\pi), a(x,y)$ & $b(\pi), b(x,y)$ & $\llbracket n,k,d\rrbracket $&$kd^2/n$ \\\hline
    3 & 4 & $1+\pi+\pi^3+\pi^4=1+xy+y^3+x$  & $1+\pi^2+\pi^{5}+\pi^9=1+(xy)^2+x^2y+y$ &  $\llbracket 24, 8, 4\rrbracket $ &5.33\\
    \hline
    3 & 5 & $1+\pi+\pi^2+\pi^7=1+xy+(xy)^2+xy^2$  & $1+\pi+\pi^{4}+\pi^{10}=1+xy+xy^4+x$ &  $\llbracket 30, 10, 4\rrbracket $ &5.33\\
    \hline
    3 & 5 & $1+\pi+\pi^3+\pi^4=1+xy+y^3+xy^4$  & $1+\pi+\pi^{3}+\pi^7=1+xy+y^3+xy^2$ &  $\llbracket 30, 6, 5\rrbracket $ &5\\
    \hline 
    3 & 7 & $1+\pi+\pi^3+\pi^{13}=1+xy+y^3+xy^6$  & $1+\pi+\pi^{4}+\pi^9=1+xy+xy^4+y^2$ &  $\llbracket 42, 12, 5\rrbracket $ &7.14\\
    \hline
    3 & 7 & $1+\pi+\pi^3+\pi^{4}=1+xy+y^3+xy^4$  & $1+\pi+\pi^{6}+\pi^{10}=1+xy+y^6+xy^3$ &  $\llbracket 42, 6, 7\rrbracket $ &7\\
    \hline
    3 & 8 & $1+\pi+\pi^2+\pi^{3}=1+xy+(xy)^2+y^3$  & $1+\pi^3+\pi^{9}+\pi^{14}=1+y^3+y+x^2y^6$ &  $\llbracket 48, 6, 8\rrbracket $ &8\\
    \hline
    3 & 8 & $1+\pi+\pi^3+\pi^{10}=1+xy+y^3+xy^2$  & $1+\pi^3+\pi^{7}+\pi^{16}=1+y^3+xy^7+x$ &  $\llbracket 48, 10, 6\rrbracket $ &7.5\\
    \hline
    4 & 5 & $1+\pi+\pi^6+\pi^{15}=1+xy+x^2y+x^3$  & $1+\pi^2+\pi^{5}+\pi^{7}=1+(xy)^2+x+x^3y^2$ &  $\llbracket 40, 14, 4\rrbracket $ &5.6\\
    \hline
    4 & 5 & $1+\pi+\pi^2+\pi^{3}=1+xy+(xy)^2+(xy)^3$  & $1+\pi+\pi^{3}+\pi^{10}=1+xy+(xy)^3+x^2$ &  $\llbracket 40, 6, 6\rrbracket $ &5.4\\
    \hline
    4 & 5 & $1+\pi+\pi^4+\pi^{5}=1+xy+y^4+x$  & $1+\pi+\pi^{4}+\pi^{9}=1+xy+y^4+xy^4$ &  $\llbracket 40,8,5\rrbracket $ &5\\
    \hline
    4 & 7 & $1+\pi+\pi^2+\pi^{4}=1+xy+(xy)^2+y^4$  & $1+\pi^2+\pi^{6}+\pi^{19}=1+(xy)^2+x^2y^6+x^3y^5$ &  $\llbracket 56,8,8\rrbracket $ &9.14\\
    \hline
    4 & 7 & $1+\pi+\pi^4+\pi^{9}=1+(xy)+y^4+xy^2$  & $1+\pi+\pi^{17}+\pi^{20}=1+xy+xy^3+y^6$ &  $\llbracket 56,14,6\rrbracket $ &9\\
    \hline
    5 & 6 & $1+\pi+\pi^2+\pi^{7}=1+xy+(xy)^2+x^2y$  & $1+\pi^3+\pi^{12}+\pi^{25}=1+(xy)^3+x^2+y$ &  $\llbracket 60,12,7\rrbracket $ &9.80\\
    \hline
    5 & 6 & $1+\pi+\pi^3+\pi^{4}=1+xy+(xy)^3+(xy)^4$  & $1+\pi^2+\pi^{11}+\pi^{18}=1+(xy)^2+xy^5+x^3$ &  $\llbracket 60,6,9\rrbracket $ &8.1\\
    \hline
    5 & 7 & $1+\pi+\pi^2+\pi^{4}=1+xy+(xy)^2+(xy)^4$  & $1+\pi+\pi^{6}+\pi^{24}=1+xy+xy^6+x^4y^3$ &  $\llbracket 70,8,9\rrbracket $ &9.26\\
    \hline

\end{tabular}}

\end{table*}
In addition to the coprime-BB codes provided so far, we present other
coprime-BB codes found by Algorithm 2 in Table \ref{cbbcodes} and \ref{cbbcodes8}.


\end{document}